\documentclass[referee]{aa}

\usepackage{amsmath,amstext}
\usepackage[T1]{fontenc}
\usepackage{hyperref}
\usepackage{graphicx}
\usepackage{txfonts}

\newcommand{\vh}{\hat{v}}
\newcommand{\Vh}{\hat{V}}
\newcommand{\ch}{\hat{c}}
\newcommand{\Xh}{\hat{X}}
\newcommand{\guu}{\hat{g}_{1U}}
\newcommand{\gul}{\hat{g}_{1L}}
\newcommand{\gdu}{\hat{g}_{2U}}
\newcommand{\gdl}{\hat{g}_{2L}}

\bibpunct{(}{)}{;}{a}{}{,}

\begin{document}

\title{Digital compensation of the side-band-rejection ratio in a fully analog 2SB sub-millimeter receiver.}
\titlerunning{Digital rejection compensation in a 2SB sub-millimeter receiver.}
\author{R. Rodriguez \inst{1,4} \and
                                R. Finger \inst{1} \and
                                F.P. Mena \inst{2} \and
                                A. Alvear \inst{1} \and
                                R. Fuentes \inst{1} \and
                                A. Khudchenko \inst{3} \and
                                R. Hesper \inst{3} \and
                                A.M. Baryshev \inst{3} \and
                                N. Reyes \inst{2} \and
                                L. Bronfman \inst{1}
                                }
\institute{
Astronomy Department, University of Chile, Camino el Observatorio 1515, Santiago, Chile.
\and
Electrical Engineering Department, University of Chile, Av. Tupper 2007, Santiago, Chile.
\and
NOVA/Kapteyn Astronomical Institute, University of Groningen, Postbus 800, 9700 AV Groningen, The Netherlands.
\and
Institute of Electricity and Electronics, Universidad Austral de Chile, General Lagos 2086, Campus Miraflores, Valdivia, Región de los Ríos, Chile }

\abstract
{In observational radio astronomy, sideband-separating receivers are preferred, particularly under high atmospheric noise, which is usually the case in the sub-millimeter range. However, obtaining a good rejection ratio between the two sidebands is difficult since, unavoidably, imbalances in the different analog components appear.}
{We describe a method to correct these imbalances without making any change in the analog part of the sideband-separating receiver, specifically, keeping the intermediate-frequency hybrid in place. This opens the possibility of implementing the method in any existing receiver.}
{(i) We have built hardware to demonstrate the validity of the method and tested it on a fully analog receiver operating between 600 and 720 GHz. (ii) We have tested the stability of calibration and performance versus time and after full resets of the receiver. (iii) We have performed an error analysis to compare the digital compensation in two configurations of analog receivers, with and without intermediate frequency (IF) hybrid.}
{(i) An average compensated sideband rejection ratio of 46~dB is obtained. (ii) Degradation of the compensated sideband rejection ratio on time and after several resets of the receiver is minimal. (iii) A receiver with an IF hybrid is more robust to systematic errors. Moreover, we have shown that the intrinsic random errors in calibration have the same impact for configuration without IF hybrid and for a configuration with IF hybrid with analog rejection ratio better than 10~dB.}
{We demonstrate that compensated rejection ratios above 40~dB are obtained even in the presence of high analog rejection. Further, we demonstrate that the method is robust allowing its use under normal operational conditions at any telescope. We also demonstrate that a full analog receiver is more robust against systematic errors. Finally, the error bars associated to the compensated rejection ratio are almost independent of whether IF hybrid is present or not.}

\keywords{2SB receiver --- Digital Back-end --- FPGA --- Sideband Rejection Ratio}

\maketitle
\section{Introduction}
\begin{figure}[t]
  \centering
        \begin{tabular}{@{}c}
     \includegraphics[width=0.40\textwidth]{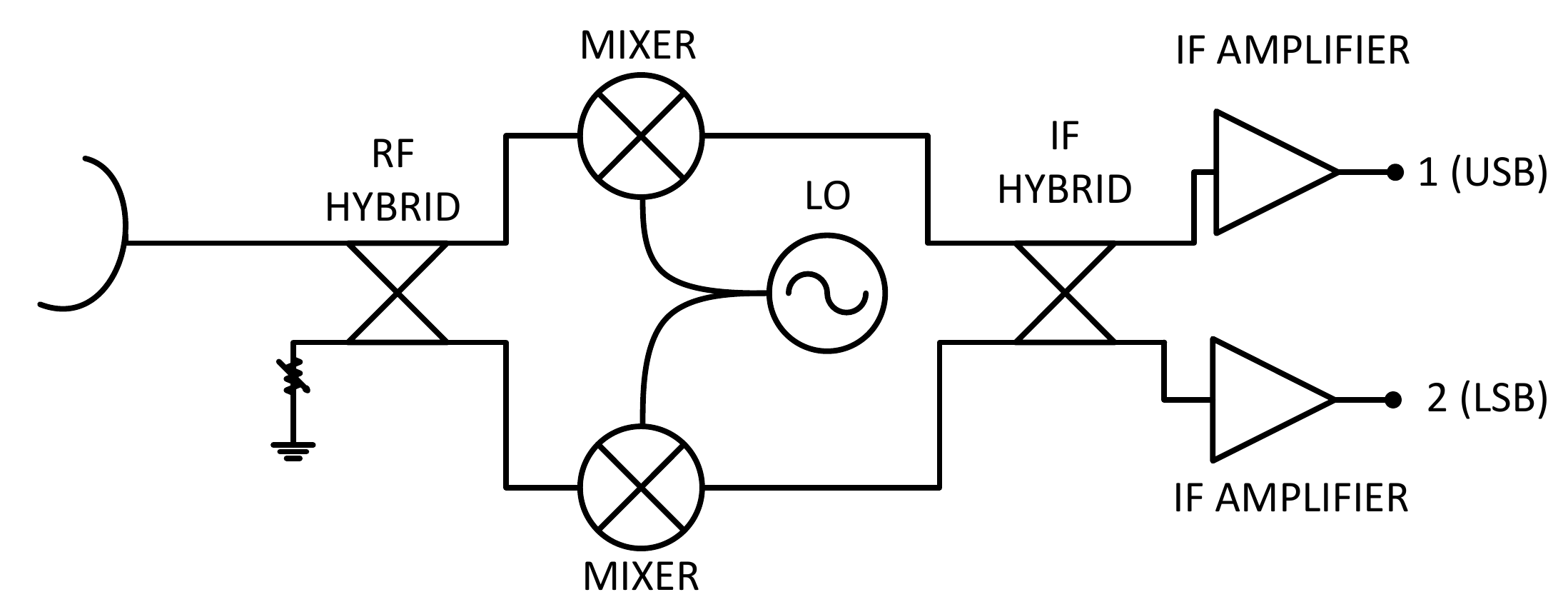} \\
     (a)\\
      \includegraphics[width=0.5\textwidth]{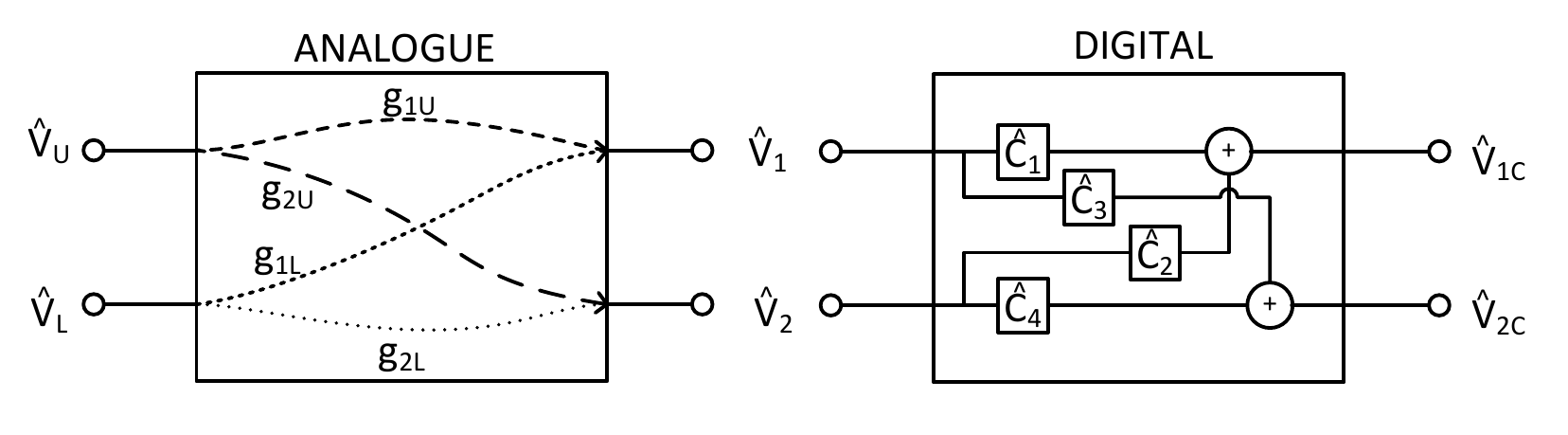} \\ 
     (b)  
    \end{tabular}
  \caption{(a) Configuration of a 2SB receiver. The RF signal is directed to a first hybrid where it is split into two arms of equal power and 90$^\circ$ phase difference. After down-conversion in two independent mixers driven by the same local oscillator (LO) signal, the resulting signals are fed into an IF hybrid. If perfect balance is achieved, the resulting signals correspond to the lower and upper sidebands. (b) Schematics of the configuration presented in this paper. An analog receiver combines the (down-converted) upper sideband and lower sideband signals with different gains while the digital part recombines them in such a way that the upper sideband and lower sideband signals are recovered at the outputs. The calibration constants $\hat{c}_i$ are determined through calibration as described in the text. We highlight that this scheme is independent of whether the analog part contains an IF hybrid or not. In other words, when no hybrid is present, IF ports~1 and~2 correspond to ports I and Q.}
  \label{fig:figure1}
\end{figure}

Radio astronomy has had an important development in recent years with the construction of some of the largest ground-based astronomical projects: \citeauthor{ALMA,SKA,FAST}, and upgrades of important radio telescopes: \citeauthor{NOEMA,VLA}. In all these projects particular effort has been made toward the improvement of front and back ends. Depending on the frequency of operation and maturity of the technology, several front-end configurations can be selected: double sideband, single sideband or sideband separating (2SB). Under non-optimal atmospheric conditions (which is the case for even the best sites at high-enough frequencies) the 2SB configuration is strongly preferred, ideally in combination with balanced mixers \citep{kerr-2ndgenB6}. This configuration, presented in Figure \ref{fig:figure1}a, is, however, more complex requiring additional components when compared to the other configurations. To achieve complete isolation between the two output ports, a perfect amplitude and phase balance along the entire radio frequency (RF) and IF chains are needed. This condition is particularly difficult to achieve over the broad RF and IF bands demanded in astronomical applications. Moreover, in classical implementations of 2SB receivers, additional RF imbalance is created by the presence of standing waves inside of the waveguide circuitry \citep{khudchenko}. Under these conditions, the sideband-rejection ratio (SRR) achieved by state-of-the-art 2SB receivers varies strongly within the RF band, usually between 7 and 30 dB \citep{band3-6,band4,band5,band7,band8}. It has been demonstrated recently that replacing the analog IF hybrid by a digital processor allows one to correct imbalances. This technique has permitted us to reach relatively constant sideban rejection ratios (SRRs) above 40~dB across the entire band in millimeter and sub-millimeter receivers \citep{AxelMurk,morgan_fisher,fisher_morgan,finger2013,rodriguez2014,spie_2014,finger2015}. Although this technique considerably improves the receiver performance, it requires a major retrofitting in existing instruments. For this reason, we demonstrate here that digital compensation of the SRR can be implemented without the necessity of removing the analog IF hybrid, making this technique readily available for upgrading existing receivers with a minimum modification on the analog front end. Subsequently, we present an exhaustive experimental analysis of the stability of the calibration, both in time and after many superconductor isolator superconductor (SIS) defluxing cycles. Furthermore, an error analysis demonstrates that, above an analog rejection ratio of 10~dB, the compensated rejection ratio has an error that is equal to the situation when no IF hybrid is present.

\section{Justification of the method}
Let us consider Figure~\ref{fig:figure1}b. The analog part of the receiver (independently of whether an IF hybrid is present or not) combines the input signals with different gains,

\begin{equation}
        \begin{aligned}
                \vh_1 &= \guu \Vh_U + \gul \Vh_L\\
                \vh_2 &= \gdu \Vh_U + \gdl \Vh_L.
        \end{aligned}
        \label{eq:v12}
\end{equation}

\noindent The digital part makes a linear combination of these voltages,

\begin{equation}
\begin{aligned}
        \vh_{1c} &= \ch_1 \vh_1 + \ch_2 \vh_2\\
        \vh_{2c} &= \ch_3 \vh_1 + \ch_4 \vh_2.
        \end{aligned}
        \label{eq:vc12}
\end{equation}

\noindent It can easily be demonstrated that, to recover the upper sideband (USB) and lower sideband (LSB) signals perfectly, the constants must satisfy

\begin{equation}
        \begin{aligned}
        \frac{\ch_1}{\ch_2} &= -\frac{\gdl}{\gul}\equiv -\Xh_2\\
        \frac{\ch_3}{\ch_4} &= -\frac{\gdu}{\guu}\equiv -\frac{1}{\Xh_1}.
        \end{aligned}
        \label{eq:c1234}
\end{equation}

The constants $\ch_1$ and $\ch_4$ can be set to $1+j0$. This particular choice will only induce an extra gain at the digital outputs. The other two constants can be obtained through calibration by injecting a well defined RF signal. In fact, it can be seen from Equations~\ref{eq:v12}, \ref{eq:vc12} and~\ref{eq:c1234} that

\begin{equation}
\begin{aligned}
\Xh_1 &= X_1 e^{j\phi_1} = \left. \frac{\vh_1}{\vh_2}\right|_{\Vh_L=0}\\
\Xh_2 &= X_2 e^{j\phi_2} = \left. \frac{\vh_2}{\vh_1}\right|_{\Vh_U=0}.
\end{aligned}
\label{eq:x12}
\end{equation}

\noindent These are the same results presented by \citet{fisher_morgan} and \citet{morgan_fisher}. However, it has to be noticed that $X$ and $\phi$ represent amplitude and phase unbalance only when no IF hybrid is present. Otherwise, $\Xh = X e^{j\phi}$ represents a complex rejection ratio. 

\section{Experiment, Results and Discussion}
\subsection{Implementation}
\begin{figure}[t]
 \centering
    \includegraphics[width=0.5\textwidth]{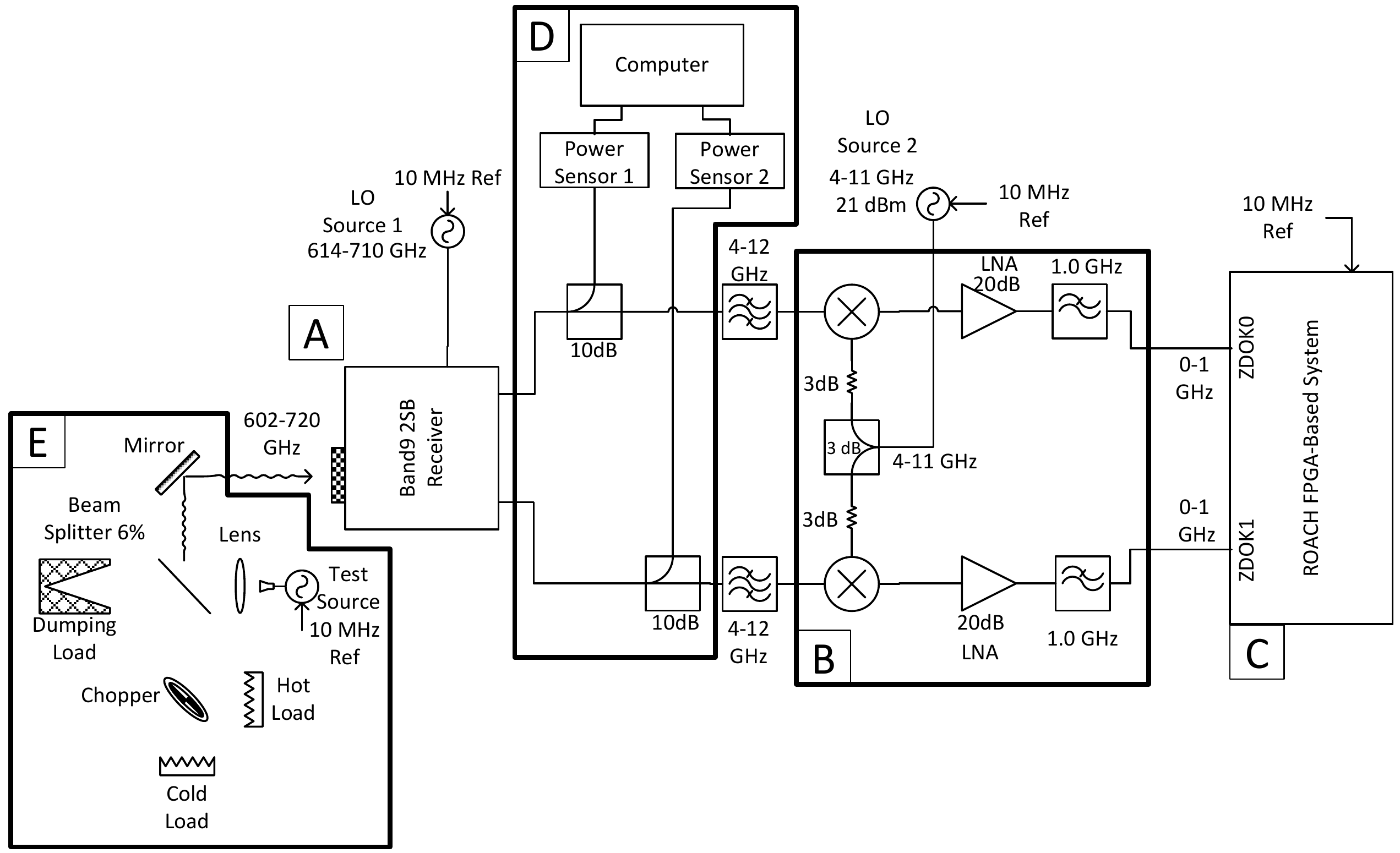}\\
    (a)\\ \vspace{2mm}
    \begin{tabular}{@{}c c@{}}
     \includegraphics[height=6.5cm]{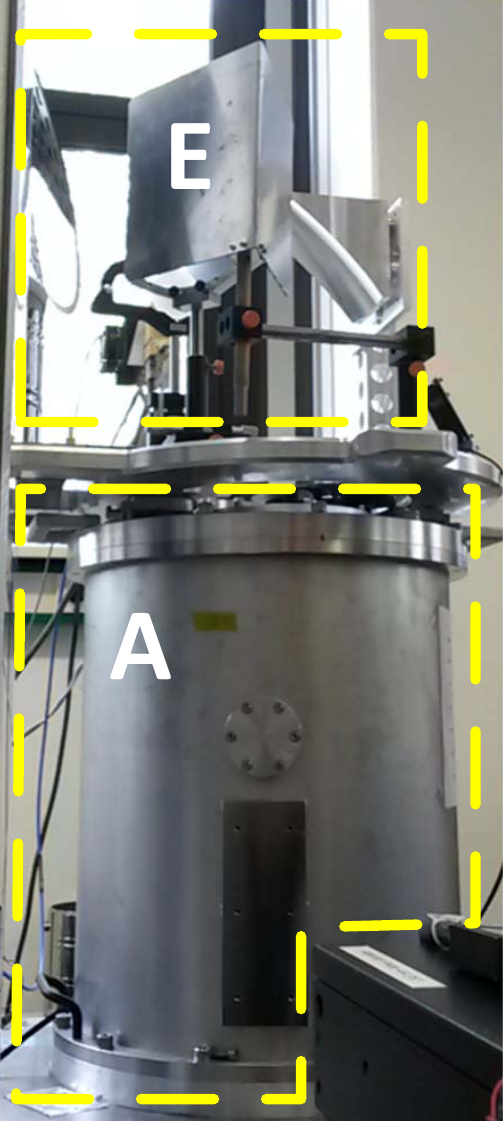} &
     \includegraphics[height=6.5cm]{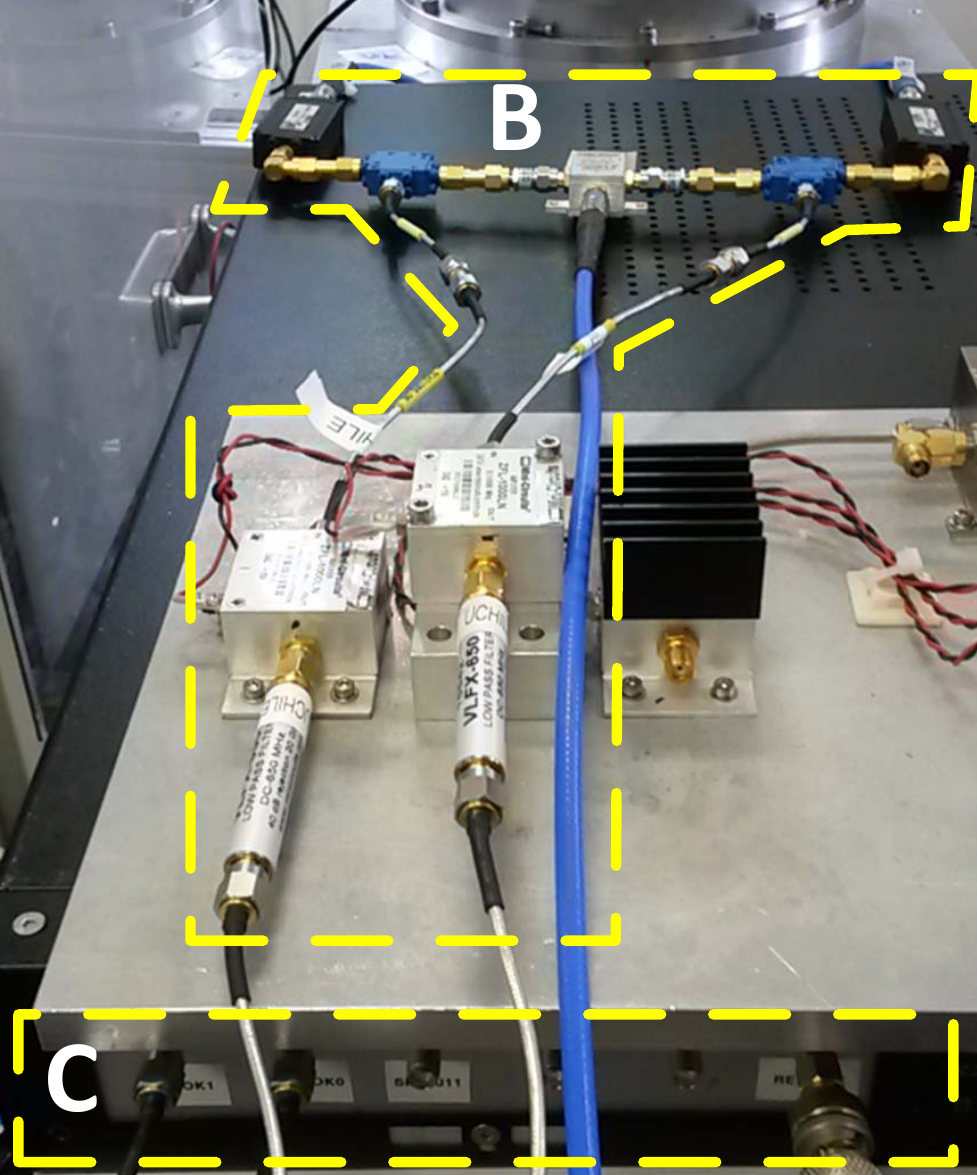}
    \end{tabular}\\
    (b)    
    \caption{(a) Schematics and (b) photographs of the experimental setup. The setup consists of a full Band-9 2SB receiver including an IF hybrid (A), a second down-conversion stage (B), and a digital back-end (C). An extra set of components, (D), is introduced to test the calibration stability with demagnetization and defluxing. This diagram also shows the RF source used for calibration and the components needed for implementing the Kerr method of measuring SRRs (E).}
    \label{fig:figure2}
\end{figure}

We have implemented the SRR-compensation scheme on a fully analog Band-9 2SB prototype receiver of ALMA \citep{hesper, khudchenko} using a FPGA-based digital spectrometer \citep{finger2013,rodriguez2014,finger2015}. This receiver uses superconductor-insulator-superconductor (SIS) junctions as mixers and operates in the RF band of  602 to 720 GHz with the standard ALMA IF frequency range,  4--12~GHz. Since the digital back-end only has 1~GHz of bandwidth, a second down-conversion is needed in order to convert the IF signal to the analog to digital converter (ADC) in the digital spectrometer. The final configuration and its implementation are presented in Figure~\ref{fig:figure2}.

\subsection{Calibration and measurement of SRRs }

\begin{figure}[t]
 \centering
    \includegraphics[width=0.5\textwidth]{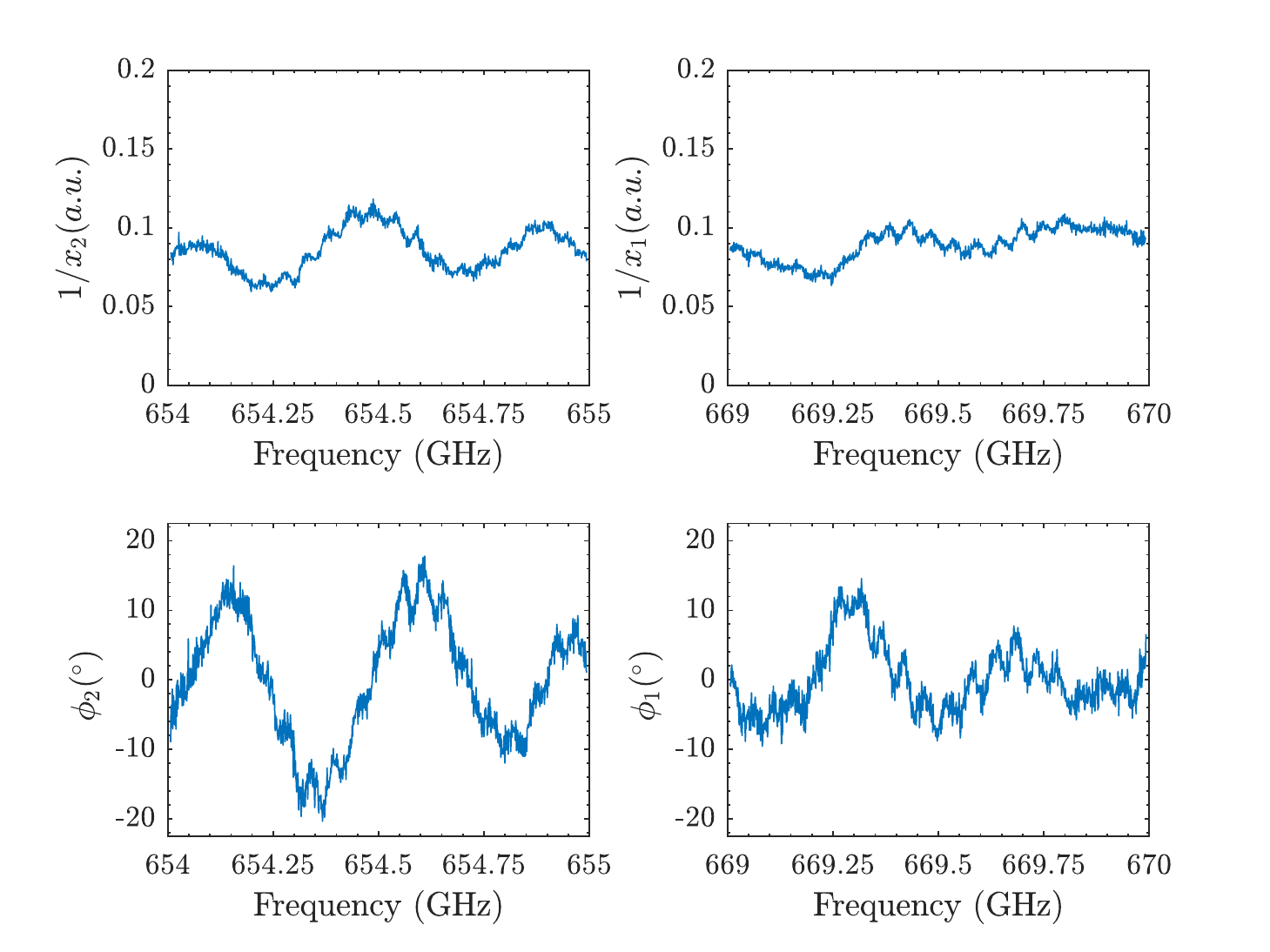}
    \caption{Example of the constants $\Xh_1$ and $\Xh_2$ measured during calibration. In this case the test tone is placed in the USB and LSB RF port for the case LO\textsubscript1 662 GHz and LO\textsubscript2 7 GHz, respectively.}
    \label{fig:figure3}
\end{figure}

The first step to compensate the SRR is to determine the values of $\Xh_1$ and $\Xh_2$. As in previous work, we have obtained them by sweeping a well known RF tone and measuring the response of the system with the digital spectrometer. Special care was taken to select a RF amplitude such that sufficient power was obtained at the output ports that, at the same time, does not saturate the SIS mixers. A typical measurement of $\Xh_1$ and $\Xh_2$ in the fully analog 2SB receiver is presented in Figure~\ref{fig:figure3}.

\begin{figure}[t]
 \centering
    \includegraphics[width=0.45\textwidth]{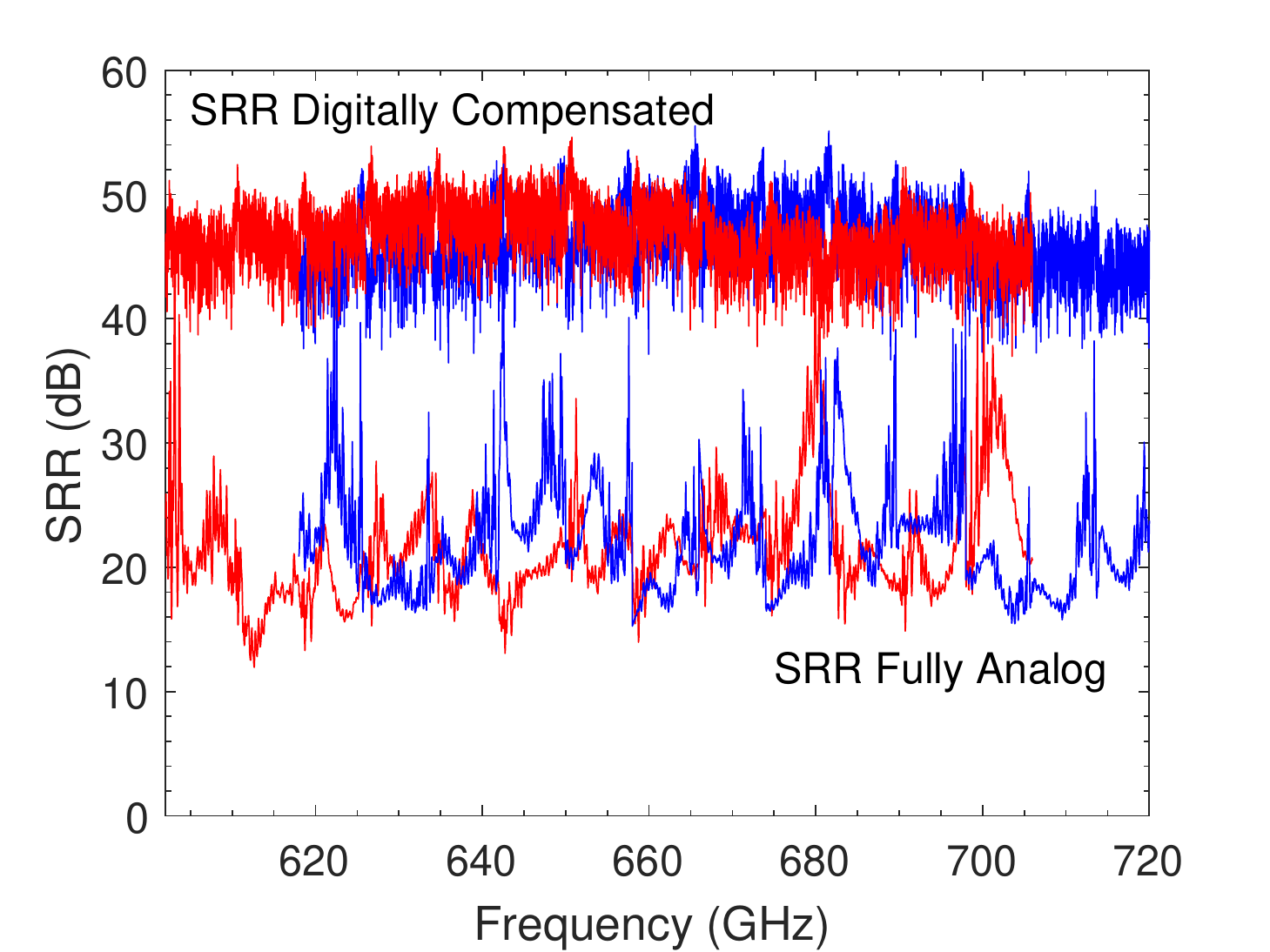}
    \caption{SRR of the 2SB receiver measured with and without digital compensation. We point out that the digitally compensated SRR has been taken at a much denser frequency grid. \textbf{Red and blue traces correspond to the LSB and USB frequencies, respectively}.  }
    \label{fig:figure4}
\end{figure}

\begin{figure}[t]
 \centering
    \begin{tabular}{@{}c}
        \includegraphics[width=0.45\textwidth]{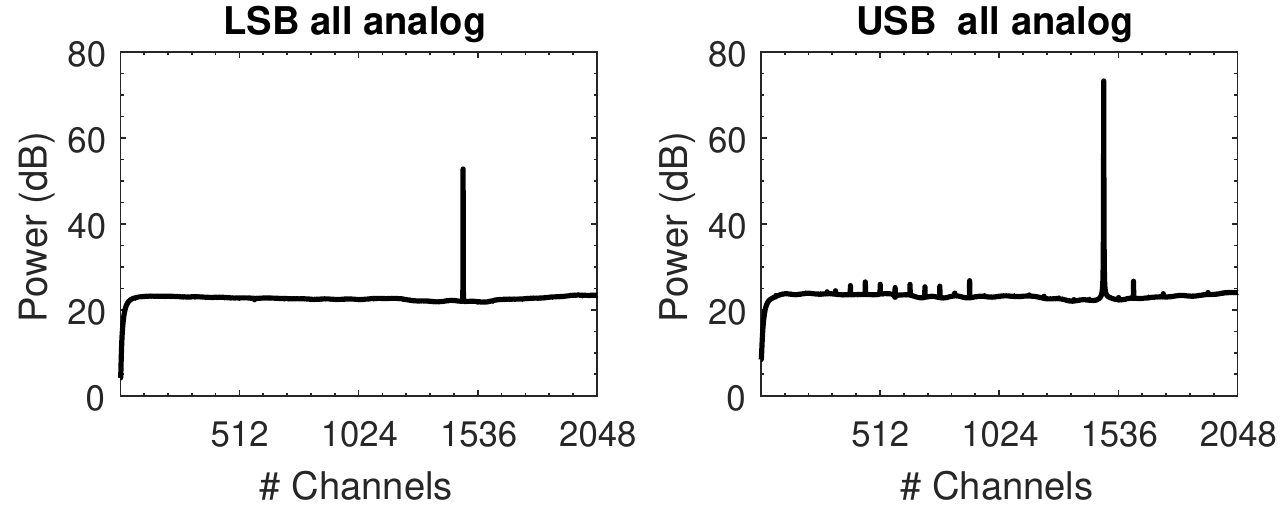} \\
        (a)\\
        \includegraphics[width=0.45\textwidth]{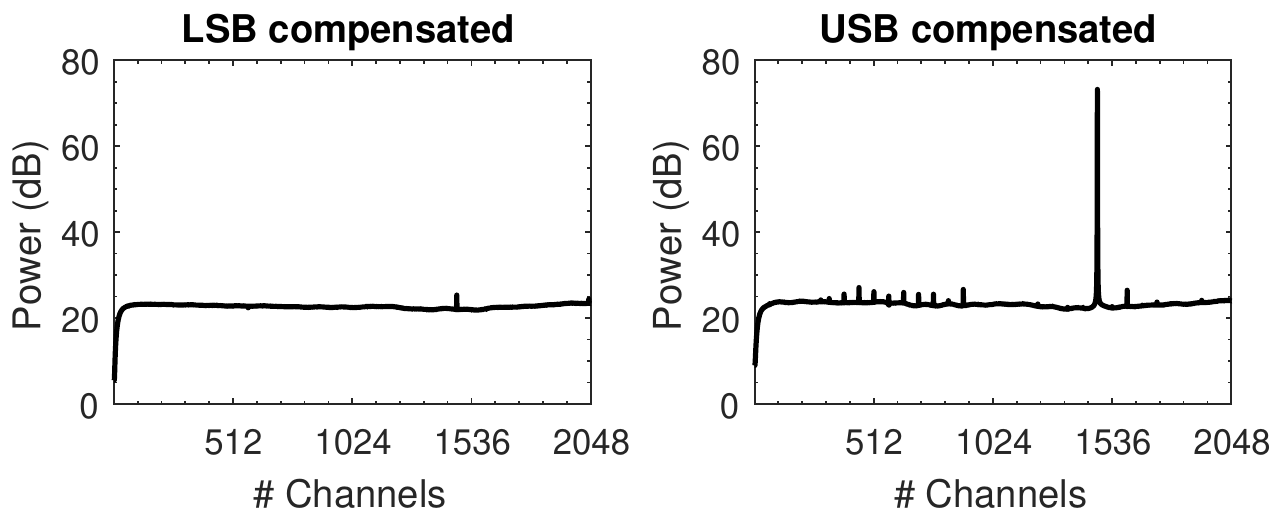} \\ 
        (b)  
    \end{tabular}
    \caption{Spectra obtained with the digital spectrometer when a RF tone of 669.7344~GHz is fed into the receiver. The tone was set at the USB with LO\textsubscript1 and LO\textsubscript2 at 662 and 7~GHz, respectively. (a) Without compensation showing a SRR of 21~dB. (b) With compensation showing a SRR of 46~dB.}
    \label{fig:figure5}
\end{figure}

The measured values of $\Xh_1$ and $\Xh_2$ allow us to obtain the four complex calibration constants needed to implement the calibration algorithm \citep{morgan_fisher}. These constants were registered in the memory of the digital spectrometer and then the SRR was measured. To measure SRR we have used the standard procedure presented by Kerr \citep{alma_memo_357}. The temperature of the loads for this test were 293 and 397~K. The tests were performed for thirteen LO\textsubscript1 and eight LO\textsubscript2 covering the full RF band. The results are presented in Figure~\ref{fig:figure4}. As a comparison, this Figure also shows the SRR measured without compensation. Furthermore, in order to illustrate the improvement when using the compensation procedure, a set of spectra is presented in Figure~\ref{fig:figure5}.

\subsection{Stability of the calibration}

For the future implementation of this technique in a telescope it is important to determine the stability of the calibration. We tested its robustness with time and resetting of the mixers. The latter is an essential procedure when SIS junctions are used as mixers. During operation, defluxing is performed regularly to eliminate trapped magnetic fluxes \cite{asayama}. In our receiver, resetting also involves demagnetization of the magnetic poles used to concentrate the magnetic field on the SIS junction \cite{hesper}. 

\begin{figure}[t]
 \centering
    \includegraphics[width=0.50\textwidth]{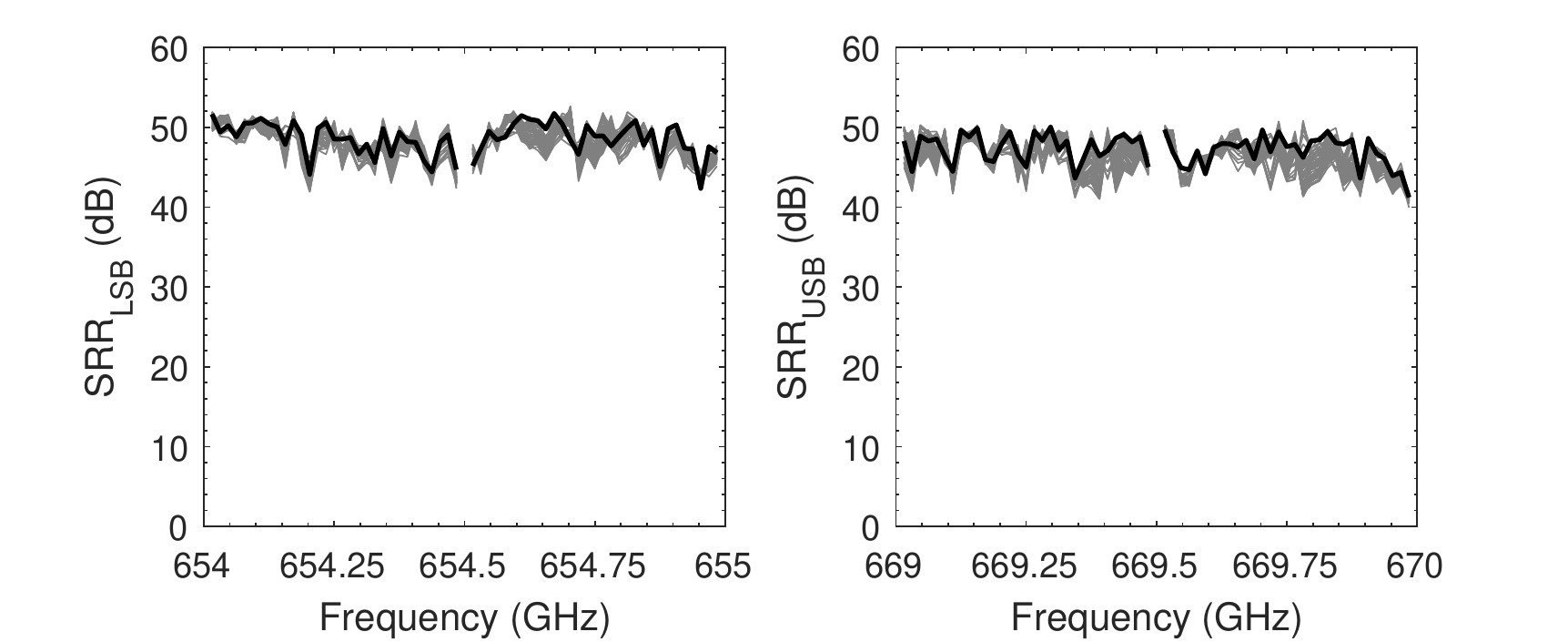}
    \caption{ Stability with time of SRR$_{LSB}$ and SRR$_{USB}$ measured at a LO\textsubscript1 of 662~GHz and LO\textsubscript2 of 7~GHz. The black line corresponds to the SRR measured immediately after calibration. Gray lines correspond to 48 consecutive measurements of the SRR, every 30 minutes, using the same initial calibration.}
    \label{fig:figure6}
\end{figure}

\begin{figure}[t]
 \centering
    \includegraphics[width=0.50\textwidth]{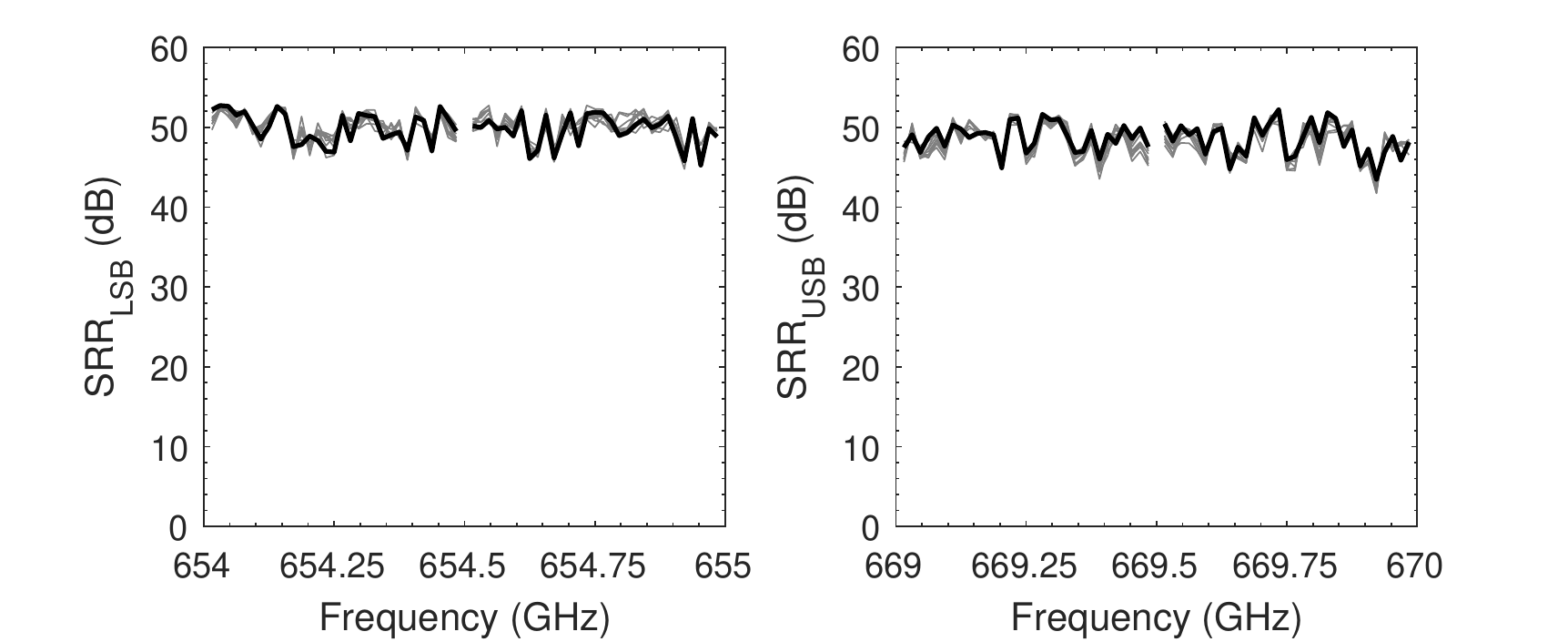}
    \caption{  Stability with defluxing of SRR$_{LSB}$ and SRR$_{USB}$ measured at a LO\textsubscript1 of 662~GHz and LO\textsubscript2 of 7~GHz. The black line corresponds to the SRR measured immediately after calibration. Gray lines correspond to nine consecutive measurements of the SRR, after defluxing the SIS, using the same initial calibration.}
    \label{fig:figure7}
\end{figure}

In the case of time stability, we first calibrated the system and then measured the SRR 48 times, once every 30 minutes. We fixed the RF such that LO\textsubscript1 was at 662 GHz and LO\textsubscript2 at 7 GHz. Figure~\ref{fig:figure6} presents the results of such measurements. For studying the stability between the application of defluxing routines, there was a small change in the setup (see Fig.~\ref{fig:figure2}). A computer, which executes an automated demagnetization and defluxing procedure, was introduced. Then, the procedure is similar to the time stability test. At the same fixed LO\textsubscript1 and LO\textsubscript2 frequencies, we first calibrated the system and measured the SRR. Afterwards, the demagnetization and defluxing routine was performed and the SRR was measured again maintaining the same calibration. We repeated this process nine times leading to the results presented in Figure~\ref{fig:figure7}. In both cases, the degradation of SRR is minimal, being 5~dB in the worst case but always above 40~dB. We highlight that degradations as small as 0.1~dB in gain and 0.5$^\circ$ in phase can produce variations of 10~dB when the SRR is in the range of 45~dB.

\subsection{Error analysis}
\subsubsection{Compensated rejection ratio}
The first step to analyze the errors when digitally compensating the rejection ratio is to express this quantity in terms of the measured quantities, the analog voltages at the output ports, $\vh_1$ and $\vh_2$. Let us consider, for example, the measurement of the USB rejection ratio (i.e., $\Vh_L=0$ in Fig.~\ref{fig:figure1}b) after calibration. Then, Equation~\ref{eq:vc12} can be rewritten as

\begin{equation}
\begin{aligned}
\vh_{1c} &= \vh_1 \left(1-\frac{1}{\Xh_{2,cal}} \frac{\vh_2}{\vh_1}\right)  \\
\vh_{2c} &= \vh_1 \left(-\frac{1}{\Xh_{1,cal}} + \frac{\vh_2}{\vh_1}\right).
\end{aligned}
\label{eq:vc12-2}
\end{equation}

\noindent There are two important points to notice in these equations. First, we have used the notation $\Xh_{i,cal}$ in order to emphasize that these constants were obtained during the calibration step. Second, the ratio $\vh_1/\vh_2$ represents the value of $\Xh_1$ taken during the measurement step. Therefore, we will use the notation $\Xh_{1,m}=\vh_1/\vh_2$. Given these considerations, the digitally compensated rejection ratio can be expressed as

\begin{equation}
        M_{Uc} = \frac{v_{1c}^2}{v_{2c}^2} = \left| \frac{\Xh_{1,cal}(\Xh_{2,cal}\Xh_{1,m}-1)}{\Xh_{2,cal}(\Xh_{1,cal}-\Xh_{1,m})}\right|^2.
        \label{eq:Muc}
\end{equation}

\noindent This equation gives an intuitive explanation of the origin of the errors of the measured compensated rejection ratio. They originate in the ability of the system to reproduce the measurement of $\Xh$. If they were equal, we would obtain perfect rejection. 

For further analysis, Equation~\ref{eq:Muc} can be simplified if we assume that $\Xh_{1,cal}=\Xh_{2,cal}$. It turns out, on one hand, that in the case of an analog receiver without IF hybrid the digitally compensated rejection ratio can be written as

\begin{equation}
M_{Uc} = \frac{1+ x^2 + 2x \cos d\phi}{ 1 +x^2 - 2x \cos d\phi}.
\label{eq:Muc1}
\end{equation}

\noindent On the other hand, for the case when the IF hybrid is present we obtain

\begin{equation}
M_{Uc} = \frac{1 + x^2 M_A^2 - 2 x M_A \cos d\phi}{M_A + x^2 M_A - 2x M_A \cos d\phi}.
\label{eq:Muc2}
\end{equation}

\noindent In both cases $x=X_{1,m}/X_{1,cal}$, $d\phi=\phi_{1,m}-\phi_{1,cal}$ and $M_A$ is the analog rejection ratio.
 
\subsubsection{Systematic errors}
\begin{figure}[t]
        \centering
        \begin{tabular}{@{}c@{}}
                \includegraphics[width=0.40\textwidth]{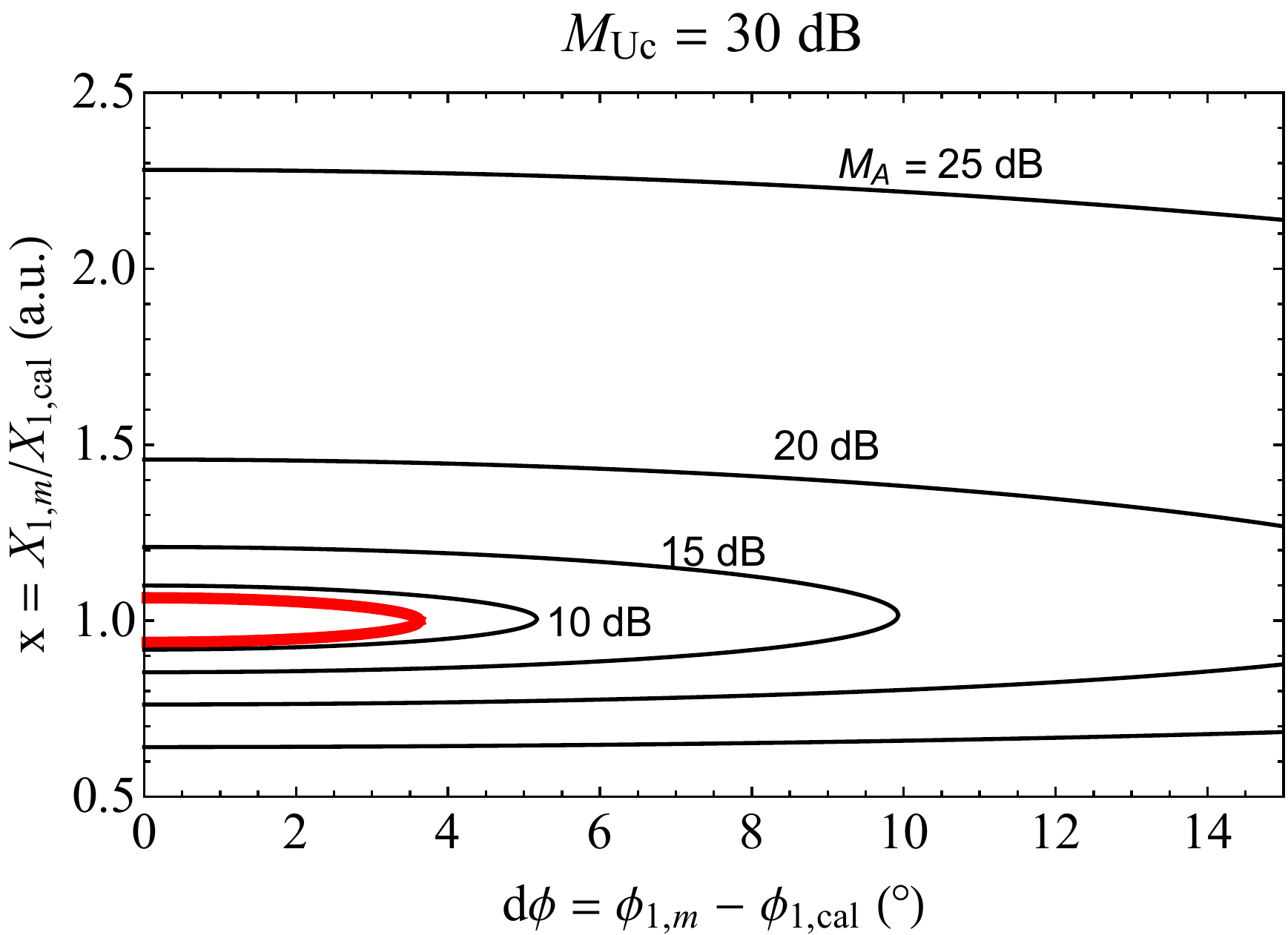} \\ (a) \\
                \includegraphics[width=0.40\textwidth]{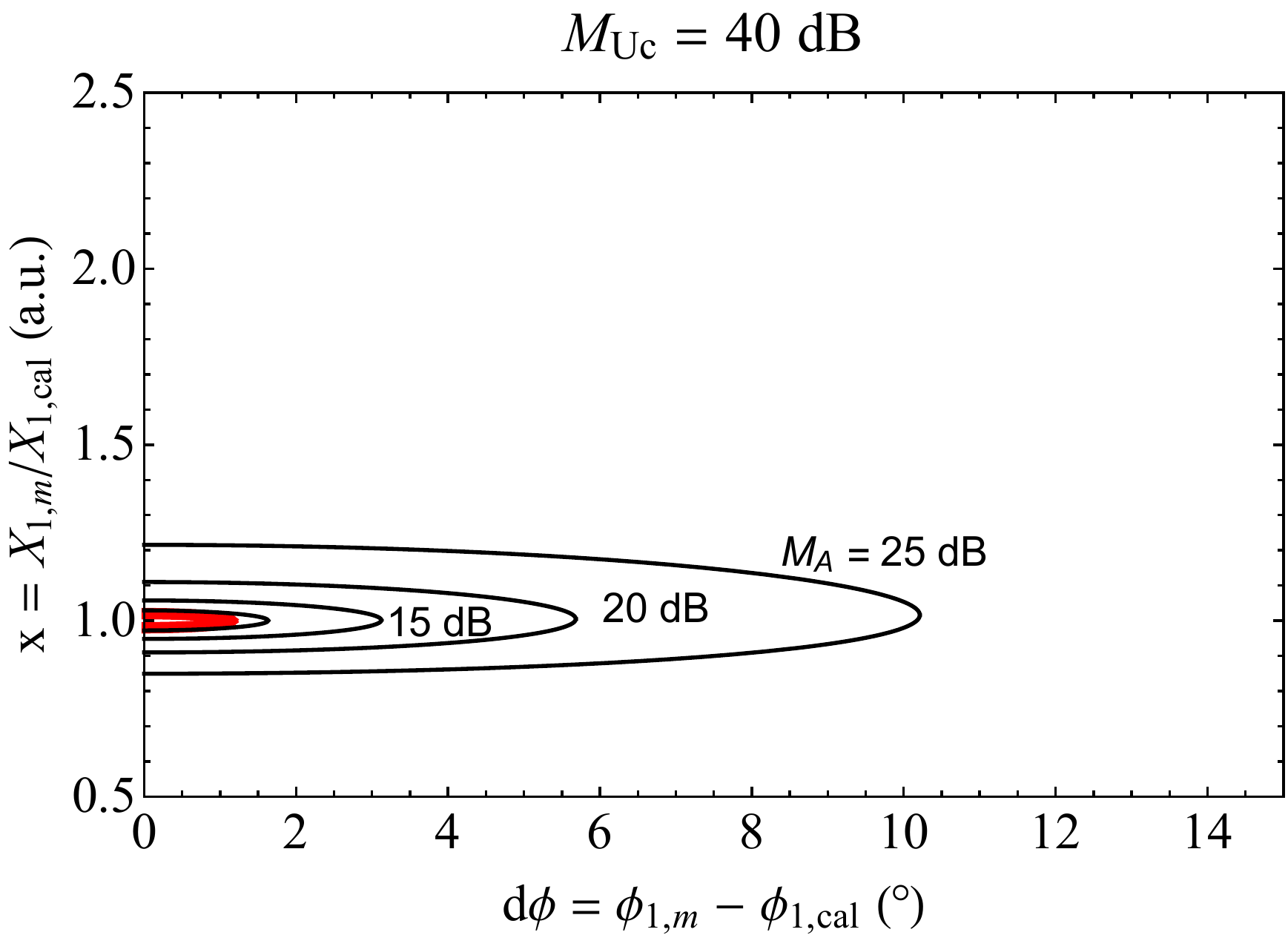} \\ (b) \\
        \end{tabular}\\
    \caption{Contour plots of the pairs $x$ and $d\phi$ that would allow to reach a desired compensated rejection ratio, $M_{Uc}$. Every contour represents a different value of analog rejection, $M_A$, of the receiver with IF hybrid. The innermost curve (thick red) represents the situation when no IF hybrid is present. (a) $M_{Uc} = 30$~dB and (b) $M_{Uc} = 40$~dB. The higher the desired compensation, the closer $\Xh_{cal}$ and $\Xh_m$ need to be.             }
    \label{fig:systematic}
\end{figure}

Systematic errors originate in additional imbalances appearing between calibration and actual measurement. \textbf{These imbalances are originated in thermal drifting or twisting of IF hybrids, for example}. Therefore, to asses them, the maximum difference between $\Xh_{cal}$ and $\Xh_m$ that can be tolerated by the system to achieve a given compensated rejection ratio should be
determined . In the model presented in the previous Section, these differences are given by $x$ and $d\phi$ of Equations~\ref{eq:Muc1} and~\ref{eq:Muc2}. Figure~\ref{fig:systematic} presents contour plots of the pairs $x$ and $d\phi$ that will produce a desired compensated rejection ratio in the presence of analog rejection ratios. The plots also include the situation when no hybrid is present. It is evident that a full analog 2SB receiver is more robust to systematic errors.

\subsubsection{Non-systematic errors}
\begin{figure}[t]
        \centering
        \begin{tabular}{@{}c c@{}}
                \includegraphics[width=0.40\textwidth]{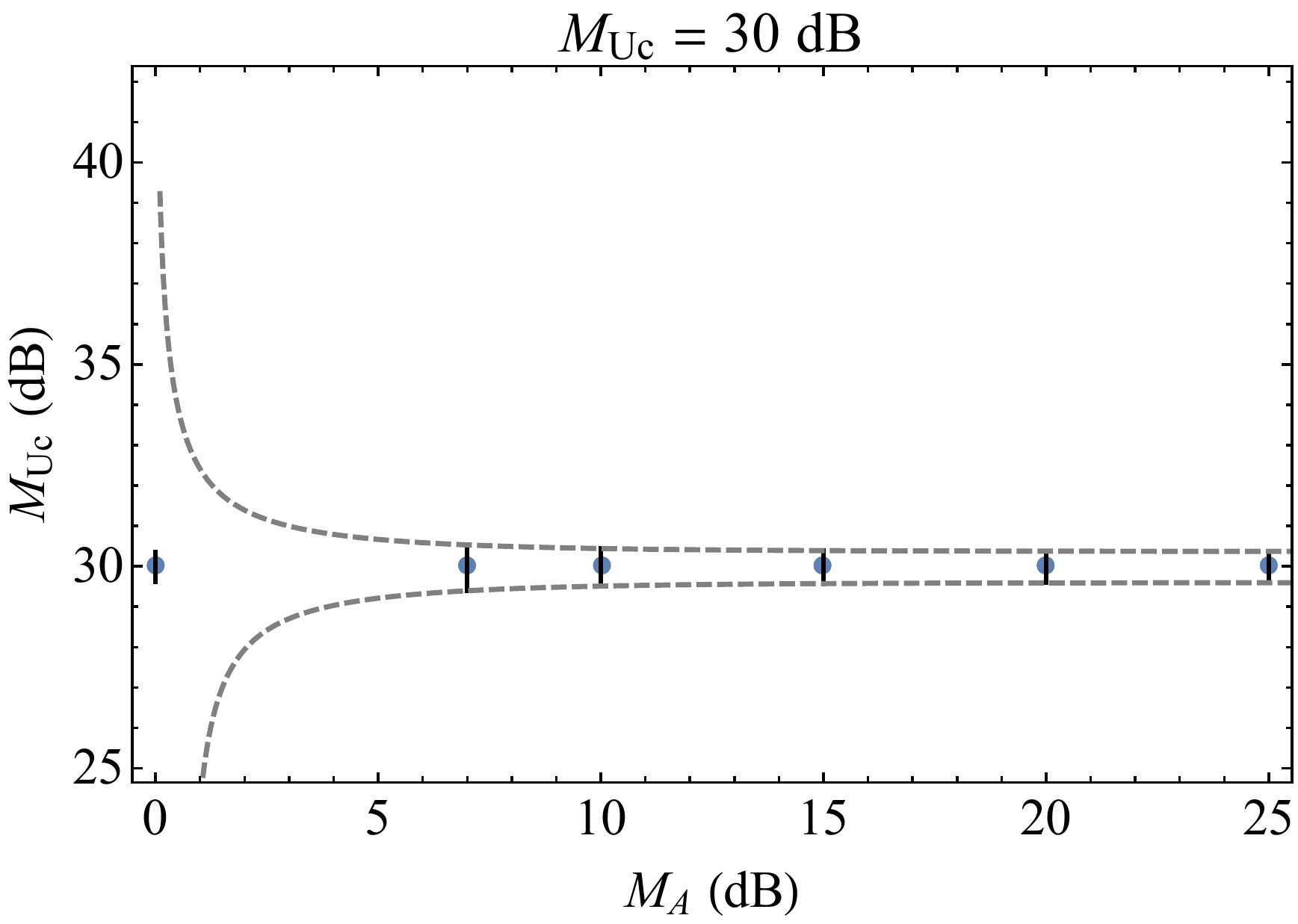} \\ (a) \\
                \includegraphics[width=0.40\textwidth]{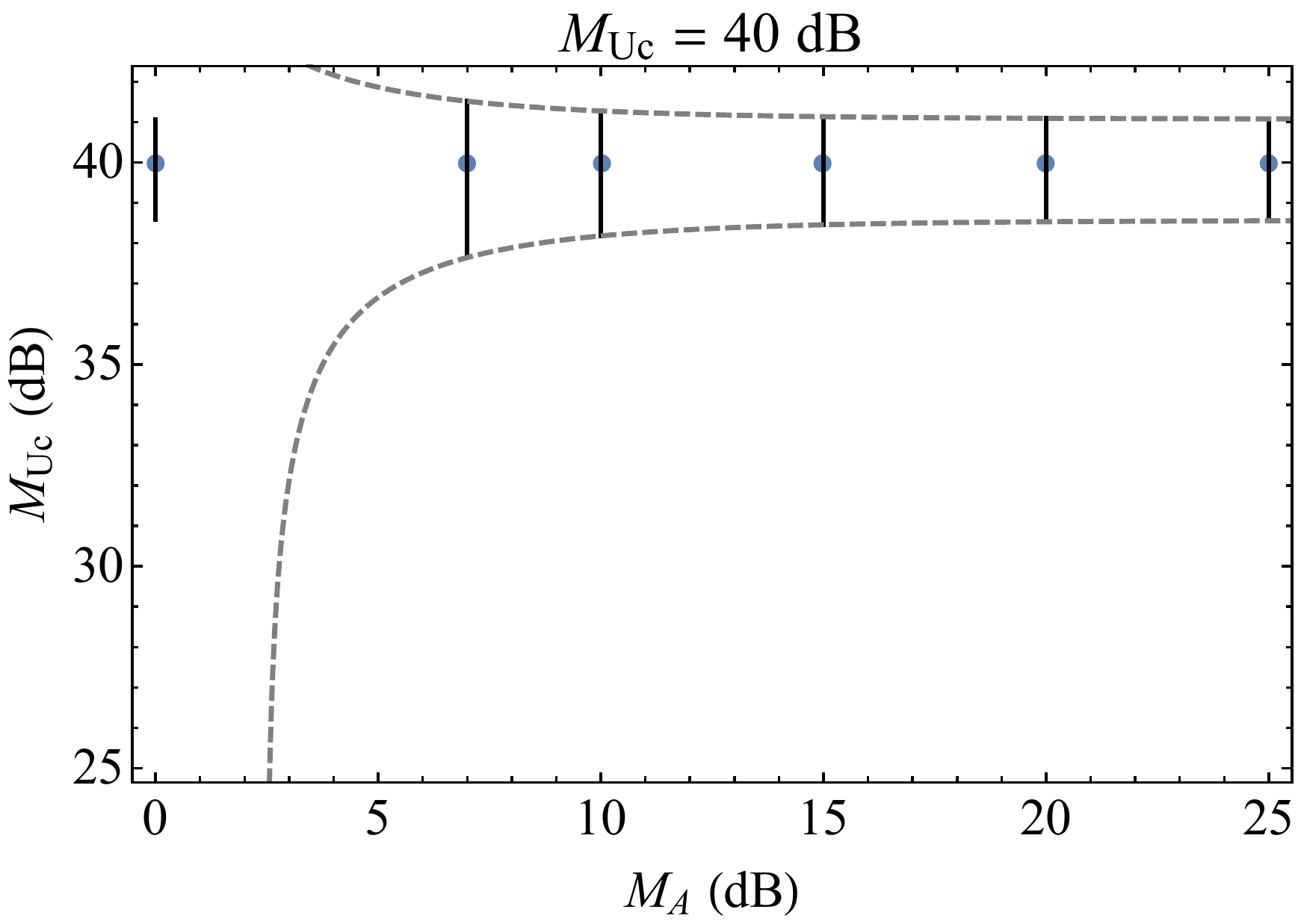} \\ (b) \\
        \end{tabular}\\
    \caption{Error analysis of the compensated rejection ratio. To construct the Figures we selected a particular $d\phi$ (in this example $d\phi=0$) and calculated the fraction $x$ that allows us to obtain a desired compensated rejection ratio, $M_{Uc}$. Then, the associated error bars were calculated. Different goal compensated rejection ratios are presented, (a) $M_{Uc} = 30$~dB and (b) $M_{Uc} = 40$~dB. Dashed gray lines are the locus of all error bars. Below a given analog rejection ratio, compensation is no longer possible. We note that, for the sake of comparison, at the point $M_A=0$ we have plotted the case where no IF hybrid is present. 
        }
    \label{fig:figure8}
\end{figure}

Random errors appear when measuring the analog voltages themselves. Basically, they are determined by the signal to noise level of the analog voltages at the output ports $\vh_1$ and $\vh_2$ or by noise of the digitizer. In order to compare the digital compensation in an analog receiver with and without an IF, we need to determine how these errors propagate to the errors in the compensated SRR. To do this, we apply the concept of propagation of errors \citep{uncertainty} to Equations~\ref{eq:Muc1} and~\ref{eq:Muc2} (see Annex). This can be done easily if we express the errors of $\Xh = X e^{j\phi}$, $\Delta X$ and $\Delta\phi$, in terms of the errors of the measured quantities, that is, the real and imaginary parts of the voltages. It can be demonstrated (see Annex) that these errors are given by

\begin{equation}
\begin{aligned}
\Delta X_i &= \frac{\Delta v}{v_i} X_i\sqrt{1+X_i^2}  \\
\Delta \phi_i &= \frac{\Delta v}{v_i} \sqrt{1+X_i^2}\;,
\end{aligned}
\label{eq:errorX}
\end{equation}

\noindent where $\Delta v$ represents the error of the real and imaginary parts of the measured voltages and $v_i$ is, in the case of a receiver with IF hybrid, the voltage in the non-rejected channel of the receiver. The last step for making a proper comparison is to consider the value of $v_i$ when the same power $P$ is coupled to the receivers. In the case of the receiver without IF hybrid we have $v_1\approx P^{1/2}/\sqrt{2}$ while for the receiver with hybrid $v_1\approx P^{1/2}\times\sqrt{M_A/(1+M_A)}$. Considering that, from our measurements in the receiver with IF hybrid, $\Delta v/v_i$ is typically of the order of $10^{-3}$ when $M_A$ is 20~dB, we have prepared Figure~\ref{fig:figure8}. This Figure summarizes the results of the error analysis outlined above and in the Annex. It demonstrates three important points: First, the error bars increase as a higher compensated rejection ratio is achieved. For the case where no hybrid is present, the error bar goes from 1.7~dB at $M_{Uc}=35$~dB to 4.9~dB at $M_{Uc}=45$~dB. Second, when the IF hybrid is present, the associated error bars in the compensated rejection ratio decrease with increasing analog rejection. Finally, above $M_A=10$~dB the error bars are practically equal to those when no IF hybrid is present.

\section{Conclusions}

We have demonstrated that digital compensation of the sideband-rejection ratio can be applied to existing all-analog 2SB receivers without the need for any modification of the front end. Specifically, we have shown that even for a fully optimized and fully analog receiver, we can improve its SRR from 22~dB to 46~dB on average. We have also shown that the technique is robust enough for typical astronomical applications since a given calibration can be used over an extended period of time and after several cycles of defluxing and demagnetization. \textbf{	 Degradation of the SRR over time, after defluxing and demagnetization is less than 5~dB in the worst case, and still maintains an SRR above 40~dB}. Furthermore, we have performed an error analysis to study how the compensation varies with and without an IF hybrid. We have demonstrated that the analog receiver with an IF hybrid is more robust to systematic errors. Regarding non-systematic errors, below an analog rejection ratio of 10~dB, the error bars in the compensated rejection ratio are larger in the fully analog 2SB receiver. Above this value the error bars in both configurations are equal. Importantly, this work demonstrates that receivers with analog rejection below 20~dB can be calibrated to reach more than 40~dB of digital sideband rejection for most of the band and that even receivers with 30~dB of analog SRR would increase its average SRR after digital compensation.

\begin{acknowledgements}
This work was supported by CONICYT through its grants CATA Basal PFB06, QUIMAL project 140002, ALMA 31150012 and Fondecyt 11140428, \textbf{ESO-Chile Joint Committee  for Development of Astronomy}, and NL NOVA ALMA R\&D project. We thank Xilinx Inc. for the donation of FPGA chips and software licenses.
\end{acknowledgements}

\section*{Annex}

The compensated rejection ratio $M_{Uc}$ is a function of the constants $\Xh_k=X_ke^{j\phi_i}$. Subsequently, the law of propagation of errors \citep{uncertainty} asserts that

\begin{equation*}
        (\Delta M_{Uc})^2=\sum_{k}
        \left[ 
        \left(\frac{\partial M_{Uc}}{\partial X_k}\right)^2(\Delta X_k)^2 +
        \left(\frac{\partial M_{Uc}}{\partial \phi_k}\right)^2(\Delta \phi_k)^2 
        \right],
\end{equation*}

\noindent where $\Delta X_k$ and $\Delta \phi_k$ are the errors associated to the measurement of the quantities $X_k$ and $\phi_k$, respectively. To evaluate the error $\Delta M_{Uc}$, the quantities $X_k$ and $\phi_k$ need to be expressed in terms of the measured voltages,

\begin{gather*}
X= \frac{v_1}{v_2} =  \sqrt { \frac{ \operatorname{Re}^2(\vh_1) +\operatorname{Im}^2(\vh_1)} {\operatorname{Re}^2(\vh_2) +\operatorname{Im}^2(\vh_2)}}  \\
\phi = \operatorname{arg}(\vh_1) - \operatorname{arg}(\vh_2) = \tan^{-1} \left[\frac{\operatorname{Re}(\vh_2) \operatorname{Im}(\vh_1) - \operatorname{Re}(\vh_1) \operatorname{Im}(\vh_2)} {\operatorname{Re}(\vh_1)\operatorname{Re}(\vh_2) + \operatorname{Im}(\vh_1)\operatorname{Im}(\vh_2)}\right],
\end{gather*}

\noindent and the law of propagation of errors should be applied to these expressions. If we assume that $\Delta\operatorname{Re}(\vh) = \Delta\operatorname{Im}(\vh) \equiv \Delta v$, Equation~\ref{eq:errorX} is obtained.

\bibliographystyle{aa}
\bibliography{referencias_articulo_b9_2SB_compensation}

\end{document}